\documentclass{aastex}
\usepackage{spr-astr-addons}
\usepackage{url}

\RequirePackage{color}

\begin{document}

\title{Correspondence between DBI-essence and Modified Chaplygin Gas
and the Generalized Second Law of Thermodynamics}

\shorttitle{Correspondence between DBI-essence and Modified
Chaplygin Gas} \shortauthors{Debnath et al.}

\author{Ujjal Debnath\altaffilmark{1}}
\and
\author{Mubasher Jamil\altaffilmark{2}}

\altaffiltext{1}{Department of Mathematics, Bengal Engineering and
Science University, Shibpur, Howrah-711 103, India. Email:
ujjaldebnath@yahoo.com , ujjal@iucaa.ernet.in}

\altaffiltext{2}{Center for Advanced Mathematics and Physics,
National University of Sciences and Technology, H-12, Islamabad,
Pakistan. Email: mjamil@camp.nust.edu.pk , jamil.camp@gmail.com}

\begin{abstract}
In this work, we have considered the DBI-essence dark energy model
in FRW Universe. We have found the exact solutions of potential,
warped brane tension and DBI scalar field. We also calculate the
statefinder parameters for our model that make it distinguishable
among numerous dark energy models. Moreover, we establish
correspondence between DBI-essence and modified Chaplygin gas (MCG)
and hence reconstruct the potential and warped brane tension. By
this reconstruction, we observe that DBI scalar field and potential
increase and warped brane tension decreases during evolution of the
Universe. Finally, we investigate the validity of the generalized
second law (GSL) of thermodynamics in the presence of DBI-essence
and modified Chaplygin gas. It is observed that the GSL breaks down
for DBI-essence model but GSL always satisfied for MCG model.
\end{abstract}


\section{Introduction}

Observations of Type Ia supernovae (SNIa) indicate that currently
the observable Universe is undergoing an accelerating expansion
\citep{riess}. This cosmic acceleration has also been confirmed by
numerous observations of large scale structure (LSS) \citep{3} and
measurements of the cosmic microwave background (CMB) anisotropy
\citep{4}. The cause for this cosmic acceleration is generally
dubbed as ``dark energy'', a mysterious exotic energy which
generates large negative pressure, whose energy density is dominated
the Universe (for a review see e.g. \citep{sami}). The astrophysical
nature of dark energy is that it does not cluster at any scale
unlike normal baryonic matter which form structures. The combined
analysis of cosmological observations suggests that the Universe is
spatially flat, and consists of about 70\% dark energy, 30\% dust
matter (cold dark matter plus baryons), and negligible radiation.
The nature of dark energy as well as its cosmological origin remain
enigmatic at present. The future of the Universe crucially depends
on the nature of dark energy: if it is quintessence (having equation
of state parameter EoS $w<-1/3$) then the energy density of
quintessence dilutes with the expansion and the acceleration will be
replaced by deceleration in far future; if the cause of cosmic
acceleration is cosmological constant ($w=-1$) then the Universe
will accelerate forever since its constant energy density provides a
continuous source of vacuum energy to produce acceleration; however
if the dark energy is phantom energy ($w<-1$) then acceleration of
the Universe will convert into super-acceleration in far future
which will eventually destroy every gravitationally stable structure
in the Universe \citep{zhang1}.

In recent years, the thermodynamics of the accelerating Universe has
got much attention and numerous interesting results are obtained. In
particular, the generalized second law (GSL) of thermodynamics has
been widely studied in the cosmological context; the law states that
the entropy of a closed isolated system along with the entropy of
its boundary is always an increasing function of time. It needs to
be stressed that this law has some informal proofs but on several
instances in cosmology, it is violated \citep{wall}. Furthermore,
the validity of GSL crucially depends on the choice of boundary of
the FRW Universe: for instance the GSL is respected if the FRW
boundary is the dynamical apparent horizon \citep{apparent} but it
is conditionally valid if the boundary is future event horizon
\citep{future}. In this work, we choose the boundary as the future
event horizon and show that the GSL is violated in all the cases
studied in this paper.

The main motivation and the organization of this work are as
follows: In section II, we introduce basic equations and solutions
for DBI-essence model. Two particular solutions are found and the
scalar field and corresponding potentials are analyzed. In section
III, we calculate statefinder parameters for DBI dark energy and its
nature are investigated during evolution of the universe. In section
IV, we develop correspondence between DBI-essence and modified
Chaplygin gas and reconstruct the potential and warped brane tension
as well as dynamics of scalar field are analyzed for two types of
solutions. In section V, we study GSL in the presence of MCG and
DBI-essence and examine the validity of GSL during evolution of the
universe bounded by the event horizon. Final section is devoted to
the discussion.

\section{\normalsize\bf{Basic equations and solutions for DBI-essence}}

Note that a simple scalar field, a `quintessence field', is a
suitable candidate as an alternative to the `cosmological constant'
\cite{q}. The dynamics of the scalar fields depend on the scalar
potentials. However, scalar fields with inverse power-law potentials
have attracted lot of research interests since in this case the
equations of motion yield attractor solutions. It ensures that the
late time behavior of the Universe is independent to the choice of
arbitrary initial conditions. However, this exquisite behavior of
the quintessence field comes at a price of extreme fine-tuning of
the cosmological parameters. This problem of fine-tuning can be
resolved if the quintessence field is modeled via approaches beyond
the standard model of particle physics, for instance, string theory
\cite{s}. In this paper, we proceed with a scalar field model where
the kinetic term is non-canonical. Such non-canonical terms in the
Lagrangian generally appear in the Braneworld gravity \cite{b}.
Here, the kinetic term has a Dirac-Born-Infeld (DBI) form.
Physically, this originates from the fact that the action of the
system is proportional to the volume traced out by the Brane during
its motion. This volume is given by the square-root of the induced
metric which automatically leads to a DBI kinetic term \cite{dbi}.

The action of the Dirac-Born-Infeld (DBI) scalar field $\phi$ can be
written as (choosing $8\pi G=c=1$) \cite{dbi}
\begin{equation}
S_{DBI}=-\int d^{4} x
\sqrt{-g}\Big[T(\phi)\sqrt{1-\frac{\dot{\phi}^{2}}{T(\phi)}}~-T(\phi)+V(\phi)
\Big],
\end{equation}
where $V(\phi)$ is the self-interacting potential and $T(\phi)$ is
the warped brane tension. In the later analysis, we shall determine
exact forms of these two functions. Now let us consider the matter
content of the Universe is composed of DBI type dark energy scalar
field. The background spacetime is the spatially flat
Friedmann-Robertson-Walker (FRW), for which the Einstein field
equations are
\begin{equation}
3H^{2}=\rho,
\end{equation}
\begin{equation}
2\dot{H}=-(\rho+p),
\end{equation}
where $H(=\frac{\dot{a}}{a})$ is the Hubble parameter. Notice that
$\dot{H}>0$ produces super-acceleration while $\dot H=0$ corresponds
to accelerated expansion due to cosmological constant.

The energy density and pressure of the scalar field are respectively
given by
\begin{equation}
\rho=\rho_{\phi}=(\gamma-1)T(\phi)+V(\phi),
\end{equation}
\begin{equation}
p=p_{\phi}=\frac{\gamma-1}{\gamma}~T(\phi)-V(\phi),
\end{equation}
where the quantity $\gamma$ is reminiscent from the usual
relativistic Lorentz factor and is given by
\begin{equation}
\gamma=\frac{1}{\sqrt{1-\frac{\dot{\phi}^{2}}{T(\phi)}}}.
\end{equation}
From above expression (6), we observe that $T(\phi)>\dot{\phi}^{2}$
and $\gamma> 1$ always. From (4) and (5) we have
$\rho_{\phi}+p_{\phi}=\frac{(\gamma^{2}-1)}{\gamma}~T(\phi)$ which
is always $>0$. From energy conservation equation, we have the wave
equation for $\phi$ as
\begin{equation}
\ddot{\phi}-\frac{3T'(\phi)}{2T(\phi)}~\dot{\phi}^{2}+
T'(\phi)+\frac{3}{\gamma^{2}}~\frac{\dot{a}}{a}~\dot{\phi}+
\frac{1}{\gamma^{3}}[V'(\phi)-T'(\phi)]=0.
\end{equation}
Now we consider two cases: (I) $\gamma=$ constant and (II)
$\gamma\ne$ constant.

\textbf{Case I:} $\gamma=$ constant. In this case, assume
$T(\phi)=n\dot{\phi}^{2}$, ($n>1$) and $V(\phi)=m\dot{\phi}^{2}$,
($m$ is a positive constant) so that $\gamma=\sqrt{\frac{n}{n-1}}$.
In these choices we have the following solutions:
\begin{equation}
a=a_{0}t^{\beta},
\end{equation}
\begin{equation}
\phi=\phi_{0}+\phi_{1}\log t,
\end{equation}
\begin{equation}
T(\phi)=n\phi_{1}^{2}e^{-\frac{2(\phi-\phi_{0})}{\phi_{1}}},
\end{equation}
and
\begin{equation}
V(\phi)=m\phi_{1}^{2}e^{-\frac{2(\phi-\phi_{0})}{\phi_{1}}},
\end{equation}
where,
\begin{eqnarray}
\beta&=&\frac{1}{3\gamma}[2(m-n)+(2n-1)\gamma^{3}],\nonumber\\
a_{0}&=&\Big(\frac{3\gamma
C\sqrt{(\gamma-1)n+m}}{\sqrt{3}[2(m-n)+(2n-1)\gamma^{3}]}\Big)^{\beta},\nonumber
\end{eqnarray}
and $\phi_{1}=\frac{C}{a_{0}}$ with $C$ and $\phi_{0}$ are
constants. From (10) and (11), we see that $T(\phi)$ and $V(\phi)$
always decrease as $\phi$ increases.

For this solution, the deceleration parameter $q$ becomes,
\begin{equation}
q=-\frac{a\ddot{a}}{\dot{a}^{2}}=-1+\frac{1}{\beta}
\end{equation}
For acceleration of the Universe, $q$ must be negative i.e.
$\beta>1$ and hence $2(m-n)+(2n-1)\gamma^{3}>3\gamma$.

\textbf{Case II}: $\gamma\ne$ constant. Let us assume,
$\gamma=\dot{\phi}^{-2}$, so from (6) we have
$T(\phi)=\frac{\dot{\phi}^{2}}{1-\dot{\phi}^{4}}>\dot{\phi}^{2}$.
Since $\gamma>1$, so that we have $\dot{\phi}^{2}<1$. Let us also
assume $V(\phi)=T(\phi)$. In this case, we have the solutions:
\begin{equation}
\dot{\phi}^{2}=\sqrt{1+\frac{1}{3\log\frac{a_{0}}{a}}},
\end{equation}
\begin{equation}
V(\phi)=T(\phi)=3\log\frac{a}{a_{0}}\times\sqrt{1+\frac{1}{3\log\frac{a_{0}}{a}}},
\end{equation}
where $a_{0}$ is the integration constant and the expression for
deceleration parameter $q$ as
\begin{equation}
q=-1-\frac{1}{2\log\frac{a_{0}}{a}},
\end{equation}
From (13) and (14), we see that the solution is valid for
$a>a_{0}e^{\frac{1}{3}}$. For acceleration of the Universe, $q$ must
be negative i.e, $a>a_{0}e^{\frac{1}{2}}$.

\section{Statefinder diagnostics for DBI-essence}

Since there are various candidates for the dark energy model, we
often face with the problem of discriminating between them, which
were solved by introducing statefinder parameters \citep{sahni}.
These statefinder diagnostic pair i.e., $\{r,s\}$ parameters are of
the form:
\begin{eqnarray}
r&=&\frac{\dddot{a}}{aH^{3}}=1+\frac{9}{2}\Big(1+\frac{p_{\phi}}
{\rho_{\phi}}\Big)\frac{\partial
p_{\phi}}{\partial\rho_{\phi}},\\
s&=&\frac{r-1}{3\Big(q-\frac{1}{2}\Big)}=\Big(1+\frac{\rho_{\phi}}
{p_{\phi}}\Big)\frac{\partial p_{\phi}}{\partial\rho_{\phi}}.
\end{eqnarray}
These parameters are dimensionless and allow us to characterize the
properties of dark energy in a model independent manner. The
statefinder is dimensionless and is constructed from the scale
factor of the Universe and its time derivatives only. The parameter
$r$ forms the next step in the hierarchy of geometrical cosmological
parameters after $H$ and $q$. For cosmological constant with a fixed
equation of state ($w=-1$) and a fixed Newton's gravitational
constant, we have $\{1,0\}$. Moreover $\{1,1\}$ represents the
standard cold dark matter model containing no radiation while
Einstein static Universe corresponds to $\{\infty,-\infty\}$
\citep{debnath23}. In literature, the diagnostic pair is analyzed
for various dark energy candidates including holographic dark energy
\citep{zhang}, agegraphic dark energy \citep{wei}, quintessence
\citep{zhang1}, dilaton dark energy \citep{dilaton}, Yang-Mills dark
energy \citep{yang}, viscous dark energy \citep{vis}, interacting
dark energy \citep{pavon}, tachyon \citep{shao}, modified Chaplygin
gas \citep{debnath1} and $f(R)$ gravity \citep{song} to name a few.

In Case I, we have the expressions of $r$ and $s$ as
$r=\frac{(\beta-1)(\beta-2)}{\beta^{2}}$ and $s=\frac{2}{3\beta}$,
which are constants.

In Case II, we have found the relation between density and pressure
as
\begin{equation}
p_{\phi}=-\rho_{\phi}+1,
\end{equation}
and the relation between $r$ and $s$ as
\begin{equation}
s=\frac{2(1-r)}{7+2r}.
\end{equation}
The behavior of the parameters $r,s$ in (19) is shown in Fig.1. From
the figure, we have seen that $s$ decreases from some negative value
to $-\infty$ as $r$ increases upto a certain stage but they obey
negative sign. After that $s$ also decreases from $+\infty$ to some
negative value as $r$ increases from negative label to positive
label during evolution of the Universe.

\begin{figure}
\includegraphics[scale=0.7]{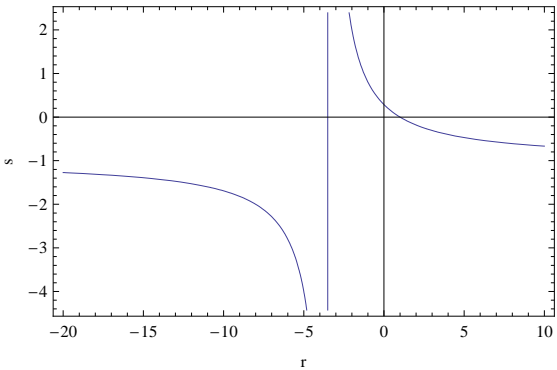}\\
\caption{It shows the Statefinder parameters $\{r,s\}$ for the DBI
model in Case II.}
\end{figure}

\section{Relation between DBI-essence and Modified Chaplygin Gas}

Here, it is interesting to find the possible relation between the
DBI-essence and the modified Chaplygin gas (MCG) \citep{ujjal}. The
MCG best fits with the $3-$year WMAP and the SDSS data with the
choice of parameters $A=-0.085$ and $\alpha=1.724$ \citep{lu} which
are improved constraints than the previous ones $-0.35<A<0.025$
\citep{jun}. Recently it is shown that the dynamical attractor for
the MCG exists at $\omega_{de}=-1$, hence MCG crosses this value
from either side $\omega_{de}>-1$ or $\omega_{de}<-1$, independent
to the choice of model parameters \citep{jing}. A generalization of
MCG is suggested in \citep{debnath} by considering $B\equiv
B(a)=B_oa^k$, where $k$ and $B_o$ are constants. The MCG is the
generalization of generalized Chaplygin gas
$p_{de}=-B/\rho_{de}^\alpha$ \citep{sen,carturan} with the addition
of a barotropic term. This special form also appears to be
consistent with the WMAP $5-$year data and henceforth the support
the unified model with dark energy and matter based on generalized
Chaplygin gas \citep{barriero,makler}. In the cosmological context,
the Chaplygin gas was first suggested as an alternative to
quintessence and demonstrated an increasing $\Lambda$ behavior for
the evolution of the Universe \citep{kamenshchik}. Recent supernovae
data also favors the two-fluid cosmological model with Chaplygin gas
and matter \citep{grigoris}. Recently, several works on Chaplygin
gas \citep{Setare1,Setare2a,Setare3a} and other dark energy model
like tachyonic field \citep{Setare2b,Setare3b} have been discussed
for interacting and non-interacting scenarios of the accelerating
universe.

In this section, we will show that, by choosing a proper potential,
the DBI-essence can be described by a modified Chaplygin gas at late
times. To find the possible relation between the DBI-essence and the
modified Chaplygin gas, we set
\begin{equation}
p_{\phi}=A\rho_{\phi}-\frac{B}{\rho_{\phi}^{\alpha}}
~~,~(A>0,~0\le\alpha\le 1).
\end{equation}
From energy conservation equation, we have the solution of
$\rho_{\phi}$ in modified Chaplygin gas as
\begin{equation}
\rho_{\phi}=\Big[\frac{B}{1+A}+\frac{C}{a^{3(1+A)(1+\alpha)}}
\Big]^{\frac{1}{1+\alpha}},
\end{equation}
where $C$ is an arbitrary positive integration constant.

From equations (4) - (6), we have
\begin{equation}
T(\phi)=\frac{\dot{\phi}^{2}(\rho_{\phi}+p_{\phi})^{2}}{(\rho_{\phi}+p_{\phi})^{2}
-\dot{\phi}^{4}},
\end{equation}
\begin{equation}
V(\phi)=\frac{\dot{\phi}^{2}\rho_{\phi}-p_{\phi}(\rho_{\phi}+p_{\phi})}{\dot{\phi}^{2}
+(\rho_{\phi}+p_{\phi})}.
\end{equation}

Now consider the following two cases:

\begin{figure}
\includegraphics[scale=0.7]{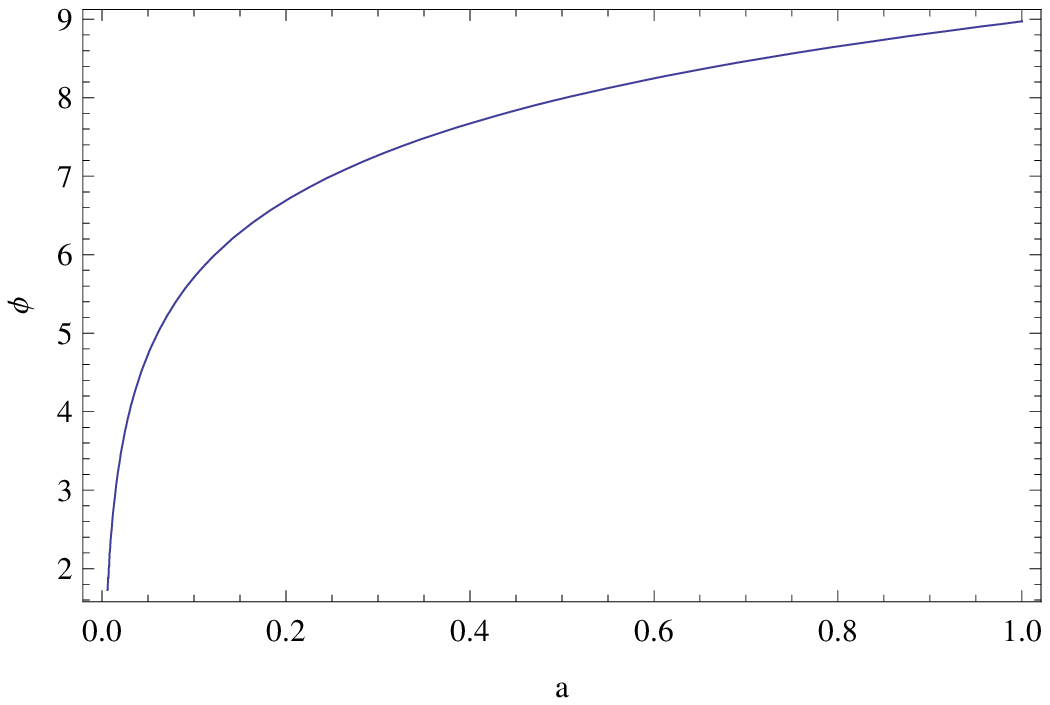}\\
\caption{It shows the variations of $\phi$ against $a$ respectively
in Case I for $A=1/3,B=0.5,C=0.5,\alpha=0.6,\gamma=2,\phi_{0}=10$.}
\end{figure}
\begin{figure}
\includegraphics[scale=0.7]{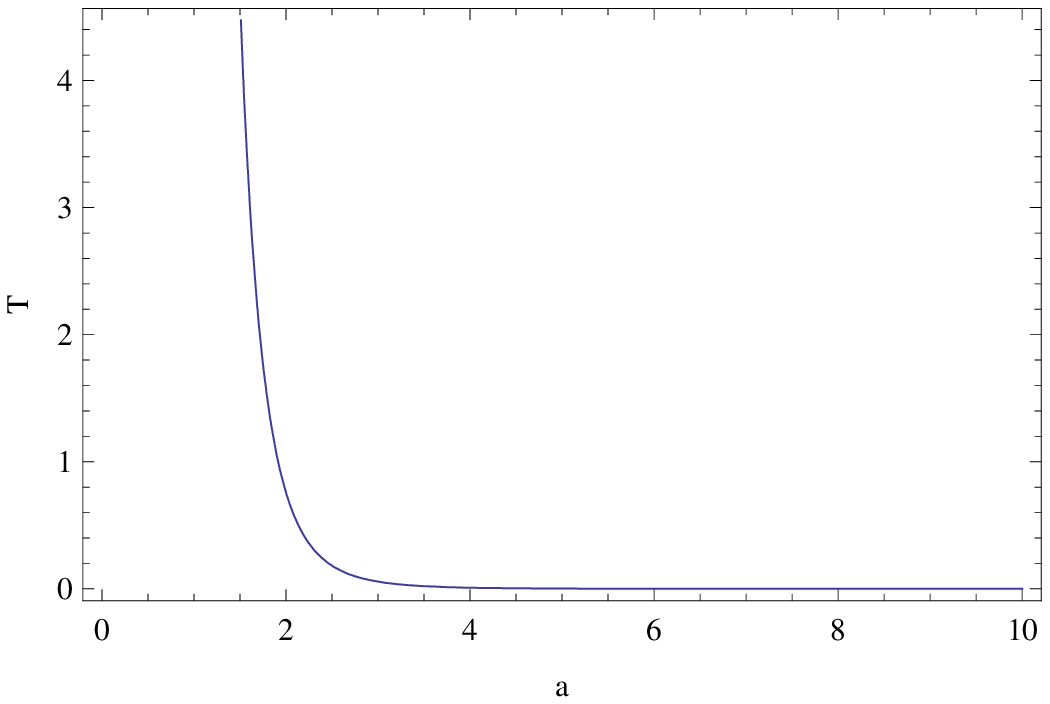}\\
\caption{It shows the variations of $T(\phi)$ against $a$
respectively in Case I for
$A=1/3,B=0.5,C=0.5,\alpha=0.6,\gamma=2,\phi_{0}=10$.}
\end{figure}
\begin{figure}
\includegraphics[scale=0.7]{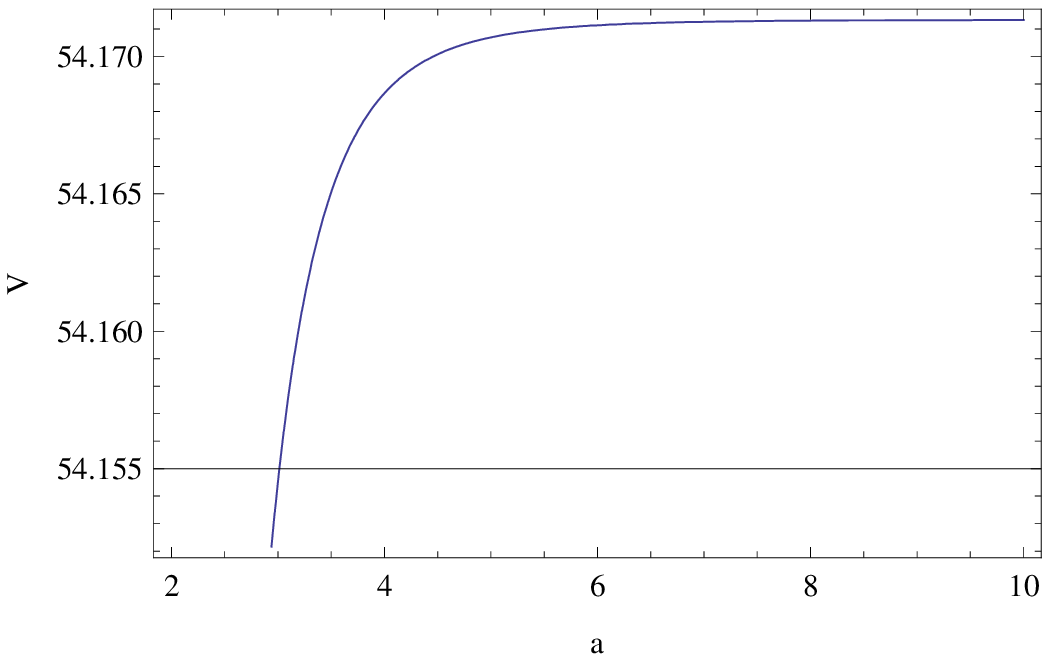}\\
\caption{It shows the variations of $V(\phi)$ against $a$
respectively in Case I for
$A=1/3,B=0.5,C=0.5,\alpha=0.6,\gamma=2,\phi_{0}=10$.}
\end{figure}

\textbf{Case I}: $\gamma=$ constant. From equations (6) and (22), it
is easy to seen that the expression of $\dot{\phi}^{2}$ is
\begin{equation}
\dot{\phi}^{2}=\frac{1}{\gamma}(\rho_{\phi}+p_{\phi}).
\end{equation}
From (20) - (24), we get the solutions
\begin{eqnarray}
\phi&=&\phi_{0}-\frac{2}{\sqrt{3\gamma(1+A)}(1+\alpha)}\nonumber\\&&\times
\tanh^{-1}\Big(\frac{\sqrt{C(1+A)+Ba^{3(1+A)(1+\alpha)}}}{\sqrt{C(1+A)}}
\Big),
\end{eqnarray}
\begin{eqnarray}
T(\phi)&=&\frac{\gamma
C(1+A)}{(\gamma^{2}-1)a^{3(1+A)}}\nonumber\\&&\times
\Big[C+\frac{B}{1+A}~a^{3(1+A)(1+\alpha)}
\Big]^{-\frac{\alpha}{1+\alpha}},
\end{eqnarray}
\begin{eqnarray}
V(\phi)&=&\frac{1}{(1+\gamma)a^{3(1+A)}}\Big[(1-\gamma
A)C+\frac{B(1+\gamma)}{1+A}~a^{3(1+A)(1+\alpha)}
\Big]\nonumber\\&&\times \Big[C+\frac{B}{1+A}~a^{3(1+A)(1+\alpha)}
\Big]^{-\frac{\alpha}{1+\alpha}}.
\end{eqnarray}
The variations of $\phi$, $T(\phi)$ and $V(\phi)$ against $a$ in
Case I have been drawn in figs. 2, 3 and 4 respectively for
$A=1/3,B=0.5,C=0.5,\alpha=0.6,\gamma=2,\phi_{0}=10$. From these
figures, we see that DBI scalar field $\phi$ and potential $V$ are
increasing and warped brane tension $T$ is decreasing with the
evolution of the Universe.

\textbf{Case II}: $\gamma\ne$ constant: Let for simplicity
$\gamma=\dot{\phi}^{-1}$. From (6) and (22) we have
\begin{equation}
\dot{\phi}^{2}=(\rho_{\phi}+p_{\phi})^{2}.
\end{equation}
The solutions are
\begin{eqnarray}
\phi&=&\phi_{0}-\frac{2}{\sqrt{3}}
a^{-\frac{3(1+A)}{2}}\nonumber\\&&\times\Big[C+\frac{B}{1+A}~a^{3(1+A)(1+\alpha)}
\Big]^{\frac{1}{2(1+\alpha)}},
\end{eqnarray}
\begin{eqnarray}
T(\phi)&=&C^{2}(1+A)^{2}\Big[C+\frac{B}{1+A}~a^{3(1+A)(1+\alpha)}
\Big]^{\frac{2}{1+\alpha}}\nonumber\\&&\times\Big(\Big[C+\frac{B}{1+A}~a^{3(1+A)(1+\alpha)}
\Big]^{2}\nonumber\\&&-C^{2}(1+A)^{2}\Big[C+\frac{B}{1+A}~a^{3(1+A)(1+\alpha)}
\Big]^{\frac{2}{1+\alpha}}\Big)^{-1},
\end{eqnarray}
\begin{eqnarray}
V(\phi)&=&\frac{C(1+A)\Big[C+\frac{B}{1+A}~a^{3(1+A)(1+\alpha)}
\Big]^{\frac{1}{1+\alpha}}
}{C(1+A)a^{3(1+A)}+\Big[C+\frac{B}{1+A}~a^{3(1+A)(1+\alpha)}
\Big]^{\frac{\alpha}{1+\alpha}}
}\nonumber\\&&-\frac{a^{3(1+A)}\Big[AC-\frac{B}{1+A}~a^{3(1+A)(1+\alpha)}
\Big]}{C(1+A)a^{3(1+A)}+\Big[C+\frac{B}{1+A}~a^{3(1+A)(1+\alpha)}
\Big]^{\frac{\alpha}{1+\alpha}} }.
\end{eqnarray}

The variations of $\phi$, $T(\phi)$ and $V(\phi)$ against $a$ in
Case II have been drawn in figs. 5, 6 and 7 respectively for
$A=1/3,B=0.5,C=0.5,\alpha=0.6,\phi_{0}=10$. From these figures, we
see that DBI scalar field $\phi$ and potential $V$ are increasing
and warped brane tension $T$ is decreasing with the evolution of the
Universe.

So the DBI scalar field, potential and warped brane tension can be
reconstructed by modified Chaplygin gas model.

\begin{figure}
\includegraphics[scale=0.7]{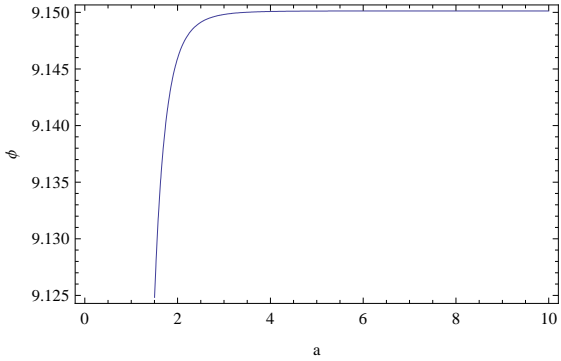}\\
\caption{It shows the variations of $\phi$ against $a$ respectively
in Case II for $A=1/3,B=0.5,C=0.5,\alpha=0.6,\phi_{0}=10$.}
\end{figure}
\begin{figure}
\includegraphics[scale=0.7]{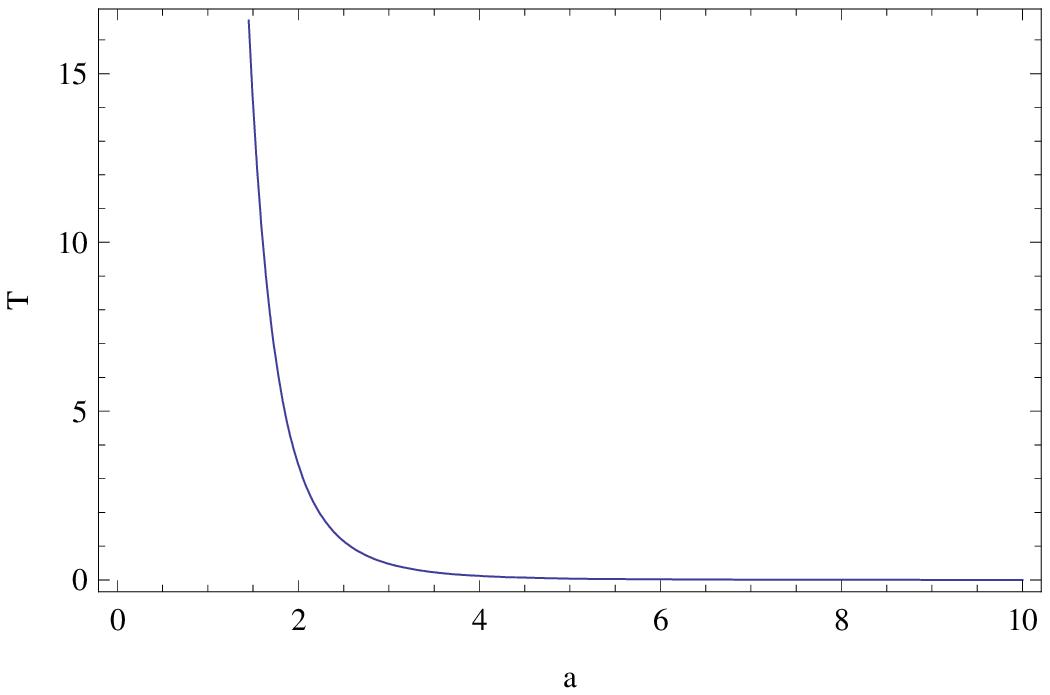}\\
\caption{It shows the variations of $T(\phi)$ against $a$
respectively in Case II for
$A=1/3,B=0.5,C=0.5,\alpha=0.6,\phi_{0}=10$.}
\end{figure}
\begin{figure}
\includegraphics[scale=0.7]{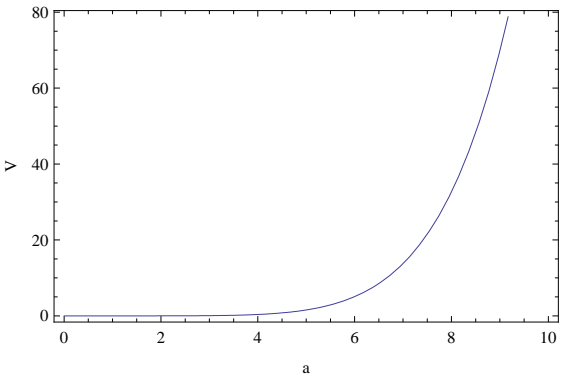}\\
\caption{It shows the variations of $V(\phi)$ against $a$
respectively in Case II for
$A=1/3,B=0.5,C=0.5,\alpha=0.6,\phi_{0}=10$.}
\end{figure}

\section{Generalized second law of thermodynamics}

Gibbons \& Hawking conjectured that event horizon area, including
cosmological event horizons, might quite generally have associated
entropy \cite{hawking}. A prominent example is de Sitter space. We
consider the FRW Universe as a thermodynamical system with the
future even horizon surface as a boundary of the system, which is a
valid assumption \cite{davis}. This horizon has got recent attention
since it yields a correct equation of state of dark energy, namely
for the holographic dark energy \cite{li}. In general, the radius of
the event horizon $R_h$ is not constant but changes with time (or
expansion of the Universe). Let $dR_h$ be an infinitesimal change in
the radius of the future event horizon during a time of interval
$dt$. This small displacement $dR_h$ will produce an infinitesimal
change $dV$ in the volume $V$ of the event horizon. Each spacetime
describing a thermodynamical system and satisfying Einstein's
equations differs infinitesimally in the extensive variables volume,
energy and entropy by $dV$, $dE$ and $dS$, respectively, while
having the same values for the intensive variables temperature $T$
and pressure $p$. Thus, for these two spacetimes describing two
thermodynamical states, there must exist some relation among these
thermodynamic quantities. To study the generalized second law of
thermodynamics through the Universe we deduce the expression for
normal entropy using the first law of thermodynamics
\begin{equation}
TdS=pdV+dE,
\end{equation}
where, $T,~S,~p,~V$ and $E$ are respectively temperature, entropy,
pressure, volume and internal energy within the event horizon. Here
the expression for internal energy can be written as $E=\rho V$. Now
the volume of the sphere is $V=\frac{4}{3}\pi R_{h}^{3}$, where
$R_{h}$ is the radius of the event horizon defined by
\begin{equation}
R_{h}=a\int_{t}^{\infty}\frac{dt}{a}=a\int_{a}^{\infty}\frac{da}{a^{2}H},
\end{equation}
which immediately gives
\begin{equation}
\dot{R}_{h}=HR_{h}-1.
\end{equation}
The temperature of the event horizon is \cite{future}
\begin{equation}
T=\frac{1}{2\pi R_{h}}.
\end{equation}
So using the above relations (33)-(35), equation (32) can be written
as
\begin{equation}
\dot{S}=\frac{8\pi \dot{H}R_{h}^{2}}{T}=16\pi^{2}\dot{H}R_{h}^{3}.
\end{equation}
Also the entropy on the event horizon is \cite{davis}
\begin{equation}
S_{h}=\frac{\pi R_{h}^{2}}{G}=8\pi^{2}R_{h}^{2}~,
\end{equation}
Using (34), (36) and (37), we have the rate of change of total
entropy as
\begin{equation}
\dot{S}+\dot{S}_{h}=16\pi^{2}R_{h}(\dot{H}R_{h}^{2}+HR_{h}-1).
\end{equation}
The generalized second law states that total entropy can not be
decrease i.e.,
\begin{equation}
\dot{S}+\dot{S}_{h}\ge 0,~~i.e.,~~\dot{H}R_{h}^{2}+HR_{h}-1 \ge 0.
\end{equation}

Now we shall examine the validity of GSL of thermodynamics for
DBI-essence and modified Chaplygin gas separately.\\

$\bullet$ {\bf DBI-essence:}\\

{\bf Case I:} $\gamma =$ constant: For the DBI solution (8), the
radius of the event horizon is

\begin{equation}
R_{h}=\frac{t}{\beta-1}~,~~\beta>1
\end{equation}

In this case, the rate of change of total entropy becomes

\begin{equation}
\dot{S}+\dot{S}_{h}=-\frac{16\pi^{2}t}{(\beta-1)^{3}}<0~~\text{for~
all}~ t.
\end{equation}

So from (41) we see that, generalized second law can not be satisfied in case I for DBI-essence model. \\

\begin{figure}
\includegraphics[scale=0.7]{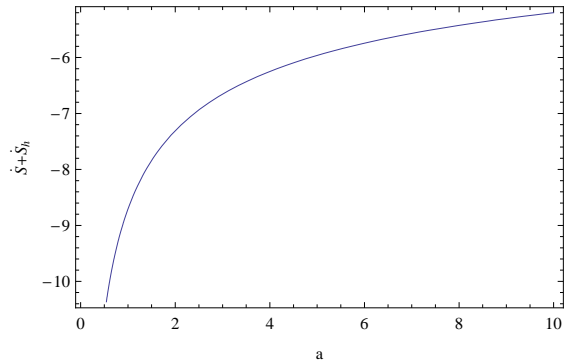}\\
\caption{It shows $\dot{S}+\dot{S}_{h}$ against $a$ in DBI model in
case II.}
\end{figure}

{\bf Case II:} $\gamma \ne$ constant: For the DBI solutions (13) and
(14) and using (2)-(6), (33) and (34), we get the radius of the
event horizon as
\begin{equation}
R_{h}=\sqrt{\pi}
~\frac{a}{a_{0}}~Erfc\Big(\sqrt{\log\frac{a}{a_{0}}}\Big),
\end{equation}
where $Erfc$ represents the complementary error function. The rate
of change of total entropy becomes
\begin{eqnarray*}
\dot{S}+\dot{S}_{h}&=&16\pi^{\frac{3}{2}}\frac{a}{a_{0}}~Erfc\Big(\sqrt{\log\frac{a}{a_{0}}}\Big)
\nonumber\\&&\times \Big[\sqrt{\pi}
~\frac{a}{a_{0}}~\sqrt{\log\frac{a}{a_{0}}}~Erfc\Big(\sqrt{\log\frac{a}{a_{0}}}\Big)-1\nonumber\\&&
+\frac{3\pi}{2}\Big(\frac{a}{a_{0}}\Big)^{2}\log\frac{a}{a_{0}}\Big(1+
\frac{1}{3\log\frac{a_{0}}{a}}
-\sqrt{1+\frac{1}{3\log\frac{a_{0}}{a}}} \Big)\nonumber\\&&\times
\Big(Erfc\Big(\sqrt{\log\frac{a}{a_{0}}}\Big) \Big)^{2} \Big]
\end{eqnarray*}
The above expression is very complicated form in $a$. So we have
drawn the figure of $\dot{S}+\dot{S}_{h}$ against $a$ in fig. 8.
From the figure, we see that $\dot{S}+\dot{S}_{h}<0$ for all values
of $a$. Hence we conclude that GSL cannot be satisfied during
evolution of the Universe in case II of DBI-essence model.
\begin{figure}
\includegraphics[scale=0.7]{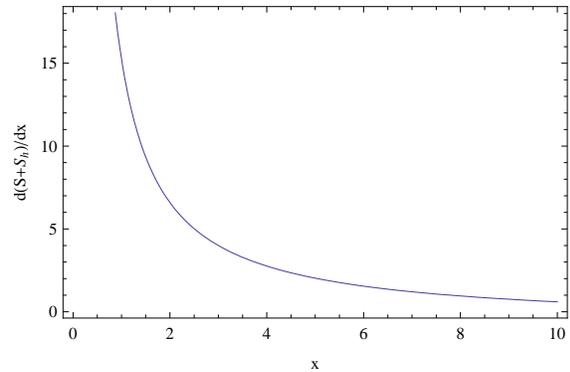}\\
\caption{Fig. 9 shows the figure of $d(S+S_{h})/dx$ against $x$ in
Modified Chaplygin gas model for $A=1/3,B=0.5,\alpha=0.6$.}
\end{figure}

$\bullet$ {\bf Modified Chaplygin Gas:}\\

For Chaplygin gas model, let us assume
$\frac{B}{1+A}~a^{3(1+A)(1+\alpha)}=Cx$, so that the solution for
density (21) reduces to
\begin{equation}
\rho=\Big(\frac{B}{1+A} \Big)^{\frac{1}{1+\alpha}}\Big(1+\frac{1}{x}
\Big)^{\frac{1}{1+\alpha}}.
\end{equation}
Using equations (2), (33) and (44), the radius of the event horizon
can be expressed as
\begin{eqnarray}
R_{h}&=&\frac{2\sqrt{3}}{1+3A}\Big(\frac{B}{1+A}
\Big)^{\frac{1}{2(1+\alpha)}} x^{\frac{1}{3(1+A)(1+\alpha)}}
\nonumber\\&&\times\Big[D-x^{\frac{1+3A}{6(1+A)(1+\alpha)}}~
_{2}F_{1}[\frac{1+3A}{6(1+A)(1+\alpha)},\frac{1}{2(1+\alpha)},\nonumber\\&&
1+\frac{1+3A}{6(1+A)(1+\alpha)} ,-x]\Big],
\end{eqnarray}
where
\begin{eqnarray}
D&=&\Big[x^{\frac{1+3A}{6(1+A)(1+\alpha)}}
_{2}F_{1}[\frac{1+3A}{6(1+A)(1+\alpha)}\nonumber\\&&,\frac{1}{2(1+\alpha)},1+\frac{1+3A}{6(1+A)(1+\alpha)}
,-x]\Big]_{x=\infty}.
\end{eqnarray}
From equation (38), we have the deviation of total entropy as
\begin{eqnarray}
\frac{d(S+S_{h})}{dx}&=&\frac{32\pi^{2}x^{\frac{2}{3(1+A)(1+\alpha)}-1}}{(1+\alpha)(1+3A)^{2}}
\Big(\frac{B}{1+A}
\Big)^{\frac{1}{1+\alpha}}\nonumber\\&&
 \times\Big[D-x^{\frac{1+3A}{6(1+A)(1+\alpha)}}~
_{2}F_{1}[\frac{1+3A}{6(1+A)(1+\alpha)}\nonumber\\&&,\frac{1}{2(1+\alpha)},1+\frac{1+3A}{6(1+A)(1+\alpha)}
,-x]\Big]\nonumber\\&& \times \Big[
x^{\frac{1+3A}{6(1+A)(1+\alpha)}}\Big(\frac{2}{1+A}-\frac{3}{1+x}
\Big)\nonumber\\&&\times_{2}F_{1}[\frac{1+3A}{6(1+A)(1+\alpha)},\frac{1}{2(1+\alpha)},\nonumber\\&&
1+\frac{1+3A}{6(1+A)(1+\alpha)} ,-x]\nonumber\\&&
+\Big(\frac{1}{1+A}+\frac{3}{1+x} \Big)D\nonumber\\&&
-\frac{1+3A}{1+A} x^{\frac{1+3A}{6(1+A)(1+\alpha)}}
(1+x)^{-\frac{1}{2(1+\alpha)}} \Big].
\end{eqnarray}
The above expression is very complicated form in $x$. So we have
drawn the figure of $d(S+S_{h})/dx$ against $x$ in fig. 9. From the
figure, we see that $d(S+S_{h})/dx>0$ for all values of $x$. Hence
we conclude that GSL always satisfied during evolution of the
Universe in modified Chaplygin gas model.

\section{Concluding Remarks}

In this work, we have considered the FRW Universe with DBI-essence
dark energy model, which is a scalar field having a non-canonical
kinetic term. We have found the exact solutions of potential, warped
brane tension and DBI scalar field. We also calculate the
statefinder parameters for our model that make it distinguishable
among numerous dark energy models. From the fig. 1, we have seen
that $s$ decreases from some negative value to $-\infty$ as $r$
increases up to a certain stage but they obey negative sign. After
that $s$ also decreases from $+\infty$ to some negative value as $r$
increases from negative label to positive label during evolution of
the Universe. Moreover, we establish correspondence between
DBI-essence and modified Chaplygin gas and hence we have
reconstructed the potential and warped brane tension in cases I and
II. By this reconstruction and from figs. 2-7, we have seen that DBI
scalar field and potential increase and warped brane tension
decreases during evolution of the Universe.

We have also considered total entropy as sum of the entropies of a
cosmological event horizon and the entropy of the DBI-essence. We
have investigated the validity of the generalized second law (GSL)
of thermodynamics in the presence of DBI-essence and modified
Chaplygin gas separately. In all of the cases (cases I and II) of
DBI-essence model, we have observed, the time derivative of the
total entropy is remaining at the negative level during the
evolution. This means that the total entropy is a decreasing
function of time in the situations considered in this work. So the
GSL breaks down for DBI-essence model in both the cases. It is also
observed that the GSL always satisfied for MCG model.

\subsection*{Acknowledgement}
One of the authors (UD) is thankful to IUCAA, Pune, India for
providing Associateship Programme under which part of the work was
carried out.


\begin{thebibliography}{}
\bibitem[\protect\citeauthoryear{Riess et al}{1998}]{riess} Riess A. et al., 1998 Astron. J., 116, 1009
\bibitem[\protect\citeauthoryear{Tegmark et al}{2004}]{3} Tegmark M. et al., 2004 Phys. Rev. D 69, 103501
\bibitem[\protect\citeauthoryear{Bennett et al}{2003}]{4} Bennett M. et al., 2003 Astrophys. J. Suppl. 148, 1
\bibitem[\protect\citeauthoryear{Copeland et al}{2006}]{sami} Copeland E.J., Sami M., \& Tsujikawa T., 2006 Int. J. Mod. Phys D., 15, 1753
\bibitem[\protect\citeauthoryear{Zhang \& Wu}{2007}]{zhang} Zhang X., \& Wu F-Q., 2007 Phys. Rev. D 76, 023502
\bibitem[\protect\citeauthoryear{Wall}{2009}]{wall} Wall A.C., 2009 JHEP 0906, 021
\bibitem[\protect\citeauthoryear{Jamil et al}{2010}]{apparent} Jamil M., Saridakis E.N., \& Setare M.R., 2010 Phys. Rev. D 81, 023007
\bibitem[\protect\citeauthoryear{Sadjadi \& Jamil}{2010}]{future} Sadjadi H.M. \& Jamil M., 2010 EPL 92, 69001
\bibitem[\protect\citeauthoryear{Ratra \& Peebles}{1988}]{q} Ratra B. \& Peebles P.J.E., 1988 Phys. Rev. D 37, 3406
\bibitem[\protect\citeauthoryear{Brax}{2007}]{s} Brax S., arXiv:0711.2428 [hep-ph]
\bibitem[\protect\citeauthoryear{Kachru et al}{2003}]{b} Kachru S. et al., 2003 JCAP 0310, 013
\bibitem[\protect\citeauthoryear{Martin \& Yamaguchi}{2008}]{dbi} Martin J. \& Yamaguchi M., 2008 Phys. Rev. D 77, 123508
\bibitem[\protect\citeauthoryear{Sahni et al}{2003}]{sahni} Sahni V. et al., 2003 JETP Lett. 77, 201
\bibitem[\protect\citeauthoryear{Debnath}{2003}]{debnath23} Debnath U., 2008 Class. Quant. Grav. 25, 205019
\bibitem[\protect\citeauthoryear{Zhang}{2005}]{zhang1} Zhang X., 2005 Int. J. Mod. Phys. D 14, 1597
\bibitem[\protect\citeauthoryear{Wei \& Cai}{2007}]{wei} Wei H. \& Cai R.G., 2007 Phys. Lett. B 655, 1
\bibitem[\protect\citeauthoryear{Huang et al}{2008}]{dilaton} Huang J.Z et al., 2008 Astrophys. Space Sci. 315, 175
\bibitem[\protect\citeauthoryear{Zhao}{2008}]{yang} Zhao W., 2008 Int. J. Mod. Phys. D 17, 1245
\bibitem[\protect\citeauthoryear{Zimdahl \& Pavon}{2004}]{pavon} Zimdahl W. \& Pavon D., 2004 Gen. Relativ. Grav. 36, 1483
\bibitem[\protect\citeauthoryear{Hu \& Meng}{2006}]{vis} Hu M. \& Meng X.H., 2006 Phys. Lett. B 635, 186
\bibitem[\protect\citeauthoryear{Shao \& Gui}{2008}]{shao} Shao Y. \& Gui Y., 2008 Mod. Phys. Lett. A 23, 65
\bibitem[\protect\citeauthoryear{Charaborty \& Debnath}{2007}]{debnath1} Chakraborty W. \& Debnath U., 2007 Mod. Phys. Lett. A 22, 1805
\bibitem[\protect\citeauthoryear{Setare \& Jamil}{2011}]{song} Setare M.R. \& Jamil M., 2011 Gen. Rel. Grav. 43, 293
\bibitem[\protect\citeauthoryear{Benaoum}{2002}]{ujjal} Benaoum H.B., hep-th/0205140
\bibitem[\protect\citeauthoryear{Lu}{2008}]{lu} Lu J. et al., 2008 Phys. Lett. B 662, 87
\bibitem[\protect\citeauthoryear{Dao-Jun \& Zhou}{2005}]{jun} Dao-Jun L. \& L. Xin-Zhou., 2005 Chin. Phys. Lett. 22, 1600
\bibitem[\protect\citeauthoryear{Jing et al}{2008}]{jing} Jing H. et al., Chin. Phys. Lett. 25, 347
\bibitem[\protect\citeauthoryear{Debnath}{2007}]{debnath} Debnath U., 2007 Astrophys. Space Sci. 312, 295
\bibitem[\protect\citeauthoryear{Barreiro \& Sen}{2004}]{sen} Barreiro T. \& Sen A. A., 2004 Phys. Rev. D 70, 124013
\bibitem[\protect\citeauthoryear{Carturan \& Finelli}{2003}]{carturan} Carturan D. \& Finelli F., 2003 Phys. Rev. D 68, 103501
\bibitem[\protect\citeauthoryear{Barriero}{2008}]{barriero} Barreiro et al, 2008, Phys. Rev. D 78, 043530
\bibitem[\protect\citeauthoryear{Makler et al}{2003}]{makler} Makler M. et al., 2003 Phys. Lett. B 555, 1
\bibitem[\protect\citeauthoryear{Kamenshchik et al}{2001}]{kamenshchik} Kamenshchik A. et al 2001 Phys. Lett. B 511, 265
\bibitem[\protect\citeauthoryear{Panotopoulos}{2008}]{grigoris} Panotopoulos G. 2008 Phys. Rev. D 77, 107303
\bibitem[\protect\citeauthoryear{Setare}{2007}]{Setare1} Setare M.R., 2007 Phys. Lett. B 648, 329
\bibitem[\protect\citeauthoryear{Setare}{2007a}]{Setare2a} Setare M.R., 2007 Phys. Lett. B 654, 1
\bibitem[\protect\citeauthoryear{Setare}{2007b}]{Setare3a} Setare M.R., 2007 Eur. Phys. J. C 52, 689
\bibitem[\protect\citeauthoryear{Setare}{2007c}]{Setare2b} Setare M.R., 2007 Phys. Lett. B 653, 116
\bibitem[\protect\citeauthoryear{Setare}{2009}]{Setare3b} Setare M.R., 2009 Phys. Lett. B 673, 241
\bibitem[\protect\citeauthoryear{Gibbons \& Hawking}{1977}]{hawking} Gibbons G.W. \& Hawking S.W., 1977 Phys. Rev. D 15, 2738
\bibitem[\protect\citeauthoryear{Davis et al}{2003}]{davis} Davis T.M. et al., 2003 Class. Quant. Grav. 20, 2753
\bibitem[\protect\citeauthoryear{Jamil et al}{2009}]{li} Jamil M. et al., 2009 Phys. Lett. B 679, 172








\end{thebibliography}
\end{document}